\documentclass{ifacconf}
\usepackage{graphicx}      
\usepackage{natbib}        
\usepackage{bbding}
\usepackage{amssymb}
\usepackage{amsmath}
\usepackage{xcolor}
\usepackage{lipsum}
\usepackage{comment}
\usepackage[hyphens,spaces,obeyspaces]{url}

\renewcommand{\H}{\mathcal{H}}
\newcommand{\tr}{\textrm{tr\,}}
\newcommand{\stefano}[1]{{\color{red}#1}}
\newcommand{\fps}[1]{$\blacktriangleright$#1$\blacktriangleleft$}

\begin{document}
\begin{frontmatter}  
\title{Hybrid Schrödinger-Liouville and projective dynamics} 

\author[First]{Kaja Krhac} 
\author[Second]{Frederic P. Schuller} 
\author[Third]{Stefano Stramigioli} 

\address[First]{School of Mathematics and Natural Sciences,\\
University of Wuppertal\\
Gaußstraße 20, 42119 Wuppertal, Germany}
\address[Second]{Department of Applied Mathematics,\\
University of Twente,\\ P.O. Box 217, 7500 AE Enschede, The Netherlands}
\address[Third]{Department of Electrical Engineering, RAM Lab,\\ University of Twente, \\ 7500 AE Enschede, The Netherlands}
\vspace{.5cm}
Accepted for the Proceedings of the 5th IFAC Workshop on\\ Control of Systems Governed by Partial Differential Equations 2025

\begin{abstract}               
Quantum dynamics provides the arguably most fundamental example of hybrid
dynamics: As long as no measurement takes place, the system state is  governed
by the Schrödinger-Liouville differential equation, which is however interrupted and
replaced by projective dynamics at times when measurements take place. We show
how this alternatingly continuous and projective evolution can be cast in form of
one single differential equation for a refined state space manifold and thus be made
amenable to standard port-theoretic analysis and control techniques.
\end{abstract}

\end{frontmatter}
\section{Introduction}
The axiomatic formulation of quantum mechanics takes a mathematically seemingly trivial form -- namely as a theory of linear operators on finite-dimensional complex Hilbert spaces -- if one restricts attention to quantum systems without translational degrees of freedom. The physics of this class of quantum systems (the simplest of which are those employing a two-dimensional Hilbert space and play a fundamental role as so-called qubits in cutting-edge technological applications, see \cite{Nielsen:2012yss}) is, however, as rich as the physics of those systems that require an infinite-dimensional Hilbert space. This class is thus an ideal test bed for theoretical investigations and we will use it as such.

The point we wish to address in this article is the logically impeccable, but technically patchy, description of the temporal evolution of a quantum state (described by a positive operator of unit trace) by way of two alternating dynamics, according to the axiomatic formulation of the theory: As long as no measurement takes place, the evolution of a generically mixed quantum state $\hat\rho$ (of a materially and energetically isolated quantum system) is postulated to be governed by the Schrödinger-Liouville equation
$$\dot{\hat\rho}(t) = -i[H,\hat\rho(t)]\,,$$
where $H$ is a hermitian operator that doubles as the measurement operator for the energy of the system and $[A,B]:=AB-BA$. As soon as a (non-degenerate) measurement takes place, however, this continuous evolution is interrupted in favour of an instantaneous and discontinuous change of the quantum state $\hat\rho$ to the new state that can be predicted, {\it ex ante}, to be
$$\sum_{z} P_z\hat\rho P_z\,,$$
where $z$ ranges from $1$ to the dimension of the Hilbert space and the $P_z$ are mutually orthogonal hermitian projectors uniquely determined by the spectral decomposition 
$$M=\sum_z m_z P_z$$ 
of the hermitian operator $M$ that mathematically abstracts and represents the physical measurement apparatus. Its spectrum $\{m_z\}$ encodes all possible measurement outcomes that the apparatus could display. Another {\it ex ante} prediction are the probabilities 
$$\tr(P_z \hat\rho)$$
for the different possible measurement outcomes $m_z$ to be obtained in any one particular performance of the measurement if the quantum state right before the measurement is $\hat\rho$. 

We emphasize that the above-described prediction of the discontinuous state change upon measurement is only the {\it ex ante} predictable part of the infamous collapse of the quantum state upon measurement. The full collapse, in contrast, cannot be predicted {\it ex ante}, but only be described {\it ex post}, meaning only after an actual measurement has already taken place and the actual measurement outcome $m_\textrm{\tiny obs}$ is already known. Indeed, it is postulated that, {\it ex post}, the quantum state after the measurement is best described by
$$\frac{P_\textrm{\tiny obs} \hat\rho P_\textrm{\tiny obs}}{\tr(P_\textrm{\tiny obs} \hat\rho P_\textrm{\tiny obs})}$$ than by the (however statistically compatible) {\it ex ante} prediction we laid out before. 

The technical purpose of this article is to reformulate the {\it ex ante} aspect, but not at all the {\it ex post} aspect, of the quantum measurement postulates, together with the continuous Schrödinger-Liouville evolution in the absence of measurements, in terms of one single differential equation on a refined quantum state space that combines classical and quantum degrees of freedom. 

It will turn out that a such refined framework allows for a complete absorption of the {\it ex ante} aspects of the measurement axioms into equally refined dynamics of the system and, thus, for a careful separation of the classical information about a quantum system from its quantum state. This reformulation provides the conceptual and technical foundation for a port-theoretic formulation of quantum theory including measurement (see \cite{SYS-002} and \cite{duindam2009modeling} for classical physics and \cite{krhac2023} for the case of quantum physics without measurement) and the transfer of related control techniques.    

\section{Refined quantum kinematics} 
The refined kinematics and dynamics presented in this and the following section go back to ideas and results of \cite{Diosi_2014} and \cite{Poulinconf} in the finite-dimensional case and have since been generalised to the infinite-dimensional case, in pursuit of a novel attempt to address the still elusive question of how quantum matter couples to the gravitational field; see 
\cite{Oppenheim:2022xjc},
\cite{Oppenheim:2020wdj}, \cite{Oppenheim:2018igd} and further references therein. 

Let $\H$ be a $d$-dimensional complex Hilbert space underlying the quantum aspects of a system and $Z:=\{1,\dots,d\}$ a discrete classical state space which, in the following, we will link to the possible classical measurement outcomes of any given measurement apparatus for that quantum system. 
The state space $\mathcal{S}(\H,Z)$ of a quantum-classical hybrid system of dimension $d$ is then given by the set of maps 
$$\rho: Z \to L^+_{[0,1]}(\H)$$
from a classical state space $Z$ to the set $L^+_{[0,1]}(\H)$ of positive semi-definite operators on $\H$ whose trace lies in the interval $[0,1]$, provided that
\begin{equation}\label{eqn:normcond}
    \sum_{z\in Z}\textrm{tr}(\rho(z))=1\,.
\end{equation}

In order to see in what sense these hybrid states capture both the classical and the quantum nature of the state of any measurement apparatus for a quantum system with a finite-dimensional Hilbert space, it is convenient to consider three physically relevant quantities that are induced by any given hybrid state $\rho$.
The first induced quantity is the  {\it induced quantum state} 
$$\hat\rho := \sum_{z\in Z} \rho(z)\,,$$
 which is a positive-semidefinite unit trace operator of $\H$ due to (\ref{eqn:normcond}).
The other two induced quantities are the {\it induced classical probabilities} for the classical states and the {\it potential collapsed quantum states}
$$p(z):=\tr \rho(z) \qquad\textrm{and}\qquad \hat\rho_{\textrm{collapse}}(z) :=  \frac{\rho(z)}{\tr\rho(z)}$$
associated with each classical state $z\in Z$. Note that $\hat\rho$ and all  $\hat\rho_{\textrm{collapse}}(z)$ are semi-definite and unit trace operators, while the $p(z)$ are real numbers in the interval $[0,1]$ for each $z\in Z$.
The physical interpretation of all three quantities is exhibited by the trivial identity
$$\hat\rho = \sum_{z\in Z} p(z) \hat\rho_{\textrm{collapse}}(z)\,,$$
where $p(z)$ is the probability for the measurement apparatus to show the classical state $z$, $\hat\rho_{\textrm{collapse}}(z)$ is the quantum state if the classical state is $z$ and $\hat\rho$ is the mixed state that encodes all one can predict about the measurement without reading off the classical measurement outcome. The hybrid state combines all this quantum and classical information in one quantity. 

\section{Refined quantum dynamics}
Now consider a differentiable curve $\rho: \mathbb{R} \to S(\H,Z)$ in hybrid state space. For a thus described time-dependent hybrid state, the most general completely positive linear dynamics preserving the normalisation condition (\ref{eqn:normcond}) are
given by the Diósi-Poulin equation (\cite{Diosi_2014}, \cite{Poulinconf})
\begin{multline}\label{eqn:Poulin}\frac{\partial\rho}{\partial t}(z,t) = 
\sum_{\alpha,\beta=0}^{d^2-1}\sum_{y\in Z} \big[ W^{\alpha\beta}(z,y,t) L_\alpha\rho(y,t)L_\beta^* - \\ \tfrac{1}{2} W^{\alpha\beta}(y,z,t) (L_\beta^* L_\alpha\rho(z,t)+\rho(z,t)L_\beta^* L_\alpha)\big]\,,\end{multline}
with the particular dynamics of a hybrid system depending on the choice of maps
$W^{\alpha\beta}(t): Z \times Z\to \mathbb{C}$ 
which for every time $t\in\mathbb{R}$ must satisfy the two positive-semi-definiteness  properties 
\begin{enumerate}
\item[(P1)] 
$W^{\alpha\beta}(z,z',t)$ are the components of a positive semi-definite matrix for all $z\neq z'\in Z$ 
\item[(P2)]  $W^{ab}(z,z,t)$ are the components of a positive semi-definite matrix for all $z\in Z$,
\end{enumerate}
where Greek indices range over $0,\dots,d^2-1$ and Latin indices over $1,\dots,d^2-1$ only. The operators $L_\alpha$ are chosen such as to constitute a basis for operators on the Hilbert space with the properties 
$$\langle L_\alpha, L_\beta\rangle = \delta_{\alpha\beta}\qquad\textrm{and}\qquad L_{0}=\textrm{id}_\H/\sqrt{d}\,,$$
where $\langle A,B\rangle:= \tr(A^* B)$ denotes the Hilbert-Schmidt hermitian inner product between operators. The orthonormality condition on the operators $L_\alpha$ implies, on the one hand, that one may expand any operator $A$ on $\H$ as
$$A = \sum_{\alpha=0}^{d^2-1} \langle L_\alpha,A\rangle L_\alpha = \sum_{a=1}^{d^2-1} \langle L_a,A\rangle L_a + \frac{\tr A}{d} \textrm{id}_\H$$
and, on the other hand, that the operators $L_a$ are all tracefree. 

For an infinite-dimensional Hilbert space $\mathcal{H}$ and a correspondingly continuous classical state space $Z$, the conditions (P1) and (P2) must be recast in a more sophisticated form  (\cite{Oppenheim:2018igd}), as is required, for instance, to formally underpin prior pioneering work on quantum monitoring in the refined formalism by \cite{Diosi_2014}.  

\noindent We will now see for which choice of functions $W^{\alpha\beta}$ the Diósi-Poulin equation captures an isolated finite-dimensional quantum system, and for which choice it captures a discrete classical statistical process. 
Both special cases can be combined to provide hybrid dynamics that describe  quantum dynamics on the Hilbert space and a stochastic process on the classical state space which are entirely independent. Conversely, this means that any coupling between quantum dynamics 
and a
classical stochastic process, to which we will turn in the next section in order to model measurement devices, must involve further $W^{\alpha\beta}$. 

\section{Schrödinger-Liouville dynamics}
Isolated quantum dynamics, specified by a hermitian Hamiltonian $H$, presents a special case of hybrid dynamics and is obtained by choosing
the time-independent functions
\begin{equation}\label{eqn:unitaryW}
W^{a 0}(z,z')=
W^{0a}(z,z')^*
:= -i\sqrt{d} \langle L_a,H\rangle  \delta_{zz'}
\end{equation}
for $a=1,\dots,d^2-1$ as the only non-vanishing components of the matrix $W^{\alpha\beta}(z,z')$, which is hence hermitian for all $z,z'\in Z$ and moreover satisfies the two required non-negativity conditions.
Indeed, for this choice, the Diósi-Poulin equation reduces to
\begin{multline}\frac{\partial \rho}{\partial t}(z,t) =-i\sum_{a=1}^{d^2-1}\left(\langle L_a,H\rangle L_a\rho(z) - \rho L_a^* \langle H,L_a\rangle\right)\\=-i\left[H-\frac{\tr H}{d}\textrm{id}_\H,\rho(z,t)\right]=-i[H,\rho(z,t)]\,.\end{multline}
From here it is immediate that the induced quantum state $\hat\rho(t)=\sum_{z\in Z} \rho(z,t)$ on the Hilbert space satisfies the Schrödinger equation
$$\dot{\hat\rho}(t) = -i[H,\hat\rho(t)]\,,$$ 
while the induced classical stochastic process $p(z,t) = \tr(\rho(z,t))$
on the classical state space $Z$ 
is frozen in time, since
$$\dot p(z,t) = \tr (-i[H,\rho(z,t)]) = 0$$
due to the cyclicity of the trace.

It is straightforward to see that the above unitary dynamics can be extended to yield the more general quantum master equation for ${\hat\rho}$ with Lindblad coefficients  
\begin{equation}\label{eqn:Lindbladian}
W^{ab}(z,y) = \delta_{zy} \lambda^{ab}
\end{equation}
for $a,b=1,\dots,d^2-1$, where $\delta$ is the Kronecker symbol and $\lambda^{ab}$ are the components of a positive semi-definite complex matrix, while again freezing the classical stochastic process in time. 

Note that any quantum dynamics for an isolated or closed system only employs the diagonal terms $W^{\alpha\beta}(z,z)$.  

\section{Projective measurement dynamics}
\noindent A hybrid system $(\H,Z)$ of dimension $d$ can be given dynamics whose solution converges exponentially fast to a hybrid state that yields the measurement statistics and collapse that are axiomatically postulated for a measurement device described by a hermitian operator $M$ 
and any input quantum state. 

With the aim to ultimately deduce the functions $W^{\alpha\beta}: Z\times Z\to\mathbb{C}$ that define these dynamics, we start from the spectral decomposition 
\begin{equation}\label{eqn:Mop}
M = \sum_{a=1}^d m^a P_a
\end{equation}
of the measurement operator in terms of its real spectrum $\{m^1,\dots,m^d\}$, which we assume to be non-degenerate for simplicity, and the associated mutually orthogonal hermitian projectors $P_1,\dots,P_d$. For definiteness, we agree on the physics convention to label the eigenvalues in decreasing order, so that the highest eigenvalue will be labelled $m_1$ and the lowest eigenvalue $m_d$.    

We then identify first which unique choice of functions $V^a: Z\times Z\to\mathbb{R}$ achieves that projective dynamics of the form
\begin{multline}
\dot\rho(z,t) = \sum_{y=1}^d \sum_{a=1}^d \left[V^a(z,y) P_a \rho(y,t) P_a^*\right. \\ \left.- \tfrac{1}{2}V^a(y,z) \left(P_a^* P_a \rho(z,t) + \rho(z,t) P_a^* P_a\right)\right]\label{eqn:projdyn}
\end{multline}
produce the desired exponential convergence to a hybrid state $\bar\rho$ that induces the probability distribution $\bar p(z)$ and the quantum state $\hat{\bar\rho}$ that are postulated by quantum mechanics. Note that an 
expansion of the projectors 
$$P_a = \sum_{\alpha=0}^{d^2-1} \langle L_\alpha,P_a\rangle L_\alpha$$ 
in terms of the Lindblad operators $L_\alpha$ reveals that the above projective dynamics are simply a special case of the Diósi-Poulin equation that amounts to choosing \begin{equation}\label{eqn:WfromV}
    W^{\alpha\beta}(z,y) = \sum_{a=1}^{d} V^a(z,y) \langle L_\alpha,P_a\rangle\langle P_a,L_\beta\rangle\,.
\end{equation}
This implies, in particular, that the second term within the square brackets on the right hand side of (\ref{eqn:projdyn}) ensures single-handedly that the projective dynamics preserve the normalisation condition (\ref{eqn:normcond}) in time and is forced entirely in terms of the choice of $V^a$.  

In order to identify the correct choice for the $V^a$, consider a family of different initial hybrid states $\rho^{(i,m)}$ given by
\begin{equation}\label{eqn:rhofamily}
    \rho^{(i,m)}(z) := \delta_{iz}\, P_m \qquad\textrm{for }i,m=1,\dots,d\,,
\end{equation}
which model different states of a measurement device whose `input' quantum state is the pure $M$-eigenstate $P_m$ (since indeed the quantum state induced by $\rho^{(i,m)}$ is $\hat\rho^{(i,m)}=P_m$ for all $i$), while its classical pointer is in position $i$ (since indeed the probability distribution induced by $\rho$ is $p^{(i,m)}(i)=1$ and $p^{(i,m)}(j)=0$ for all $j\neq i$). So unless $i=m$, this hybrid state corresponds to a measurement device axiomatically described by $M$ whose pointer does not yet point to the only admissible position it would have to after measurement, given the pure $M$-eigenstate $\rho^{(i,m)}=P_m$ before measurement. 

We will now show that the functions $V^a$ are uniquely determined by the requirement that the projective dynamics (\ref{eqn:projdyn}) evolve every one of these initial hybrid states $\rho^{(i,m)}$ to the hybrid state  $\bar\rho^{(i,m)}$ given by
\begin{equation}\label{eqn:barrhofamily}
\bar\rho^{(i,m)}(z)=\delta_{mz} P_m\,,
\end{equation}
which corresponds to the quantum `output' state of the measurement device being still $P_m$ (since indeed  $\hat{\bar\rho}^{(i,m)} = P_m$) while the pointer now certainly points to the position $m$ (since the probability distribution is then given by $\bar p^{(i,m)}(z)=\delta_{mz}$). 
In other words, the hybrid state $\bar\rho$ would then correspond to the quantum state and pointer position after measurement as stipulated by the quantum mechanical measurement axiom. 

Consider the projective evolution of any member $\rho^{(i,m)}$ of the family of hybrid states at the initial time, 
\begin{multline}\dot\rho^{(i,m)}(z,0) = \sum_{y=1}^{d} \sum_{a=1}^{d} \Big[V^a(z,y) P_a \rho^{(i,m)}(y) P_a^* \\ - \frac{1}{2} V^a(y,z)\left(P_a^* P_a \rho^{(i,m)}(z,0)+\rho^{(i,m)}(z,0)P_a^* P_a \right)\Big]\,,\end{multline}
which upon insertion of (\ref{eqn:rhofamily}) immediately reduces to
%
$$\dot\rho^{(i,m)}(z,0) =  V^m(z,i) P_m - \delta_{iz}\,\sum_{y=1}^d V^m(y,z) P_m\,.$$
It is now useful to divide the anaylsis into two complementary cases. 

The first case is $i=m$, in which the initial hybrid state $\rho^{(i,m)}$ is already identical to the desired final hybrid state $\bar\rho^{(i,m)}$, so that the dynamics are supposed to not change the initial state. This means that the $V$ must be chosen such as to make  
$$
\dot\rho^{(m,m)}(z,0) = \left\{ \begin{array}{ll}
 - \sum_{y\neq m} V^m(y,m) P_m & \quad z=m\\
    V^m(z,m)\,P_m & \quad z\neq m
\end{array}
\right\}
$$
vanish. Noting that the expression for $z=m$ is determined entirely by those for $z\neq m$, we see that it is necessary and sufficient that we set
\begin{equation}\label{eqn:cond1}
    V^m(z,m):=0 \qquad \textrm{for all } z\neq m\,.
\end{equation}

The second case is $i\neq m$, which requires non-trivial dynamics in order to evolve the initial hybrid state $\rho^{(i,m)}$ into the final hybrid state $\bar\rho^{(i,m)}$. For these initial states one has
$$\dot\rho^{(i,m)}(z,0) = \left\{\begin{array}{cl}
    -\sum_{y\neq i} V^m(y,i)P_m & \quad z = i \\
    V^m(m,i) P_m & \quad z\neq i \textrm{ and } z= m \\
     V^m(z,i) P_m  & \quad z\neq i \textrm{ and } z\neq m 
\end{array}\right\}.$$
Noting that the expression for $z=i$ is determined entirely by the two expressions for $z\neq i$, we can restrict attention to the latter.

To obtain the required  evolution from $\rho^{(i,m)}$ to $\bar\rho^{(i,m)}$  thus requires one to let
\begin{equation}\label{eqn:cond3}V^m(z,i) := 0 \qquad\textrm{for all } z\not\in\{m,i\}
\end{equation}
and to choose 
\begin{equation}\label{eqn:cond2}V^m(m,i) \in \mathbb{R}^+\,,
\end{equation}
both provided $m\neq i$. Note that conditions (\ref{eqn:cond1}), (\ref{eqn:cond3}), (\ref{eqn:cond2}) determine all off-diagonal terms ($z\neq y$) of $V^m(z,y)$ and thus, through (\ref{eqn:WfromV}), of $W^{\alpha\beta}(z,y)$, while it leaves all diagonal terms ($z=y$) undetermined.  

Any choice in accordance with these conditions thus presents a viable implementation of the projective measurement (\ref{eqn:Mop}) as long as the positive semi-definiteness conditions (P1) and (P2) are also satisfied. 

 


\section{An exactly solvable implementation}
The simplest dynamical implementation of a projective measurement (\ref{eqn:Mop}) is given by the choice
\begin{equation}\label{eq:VM}
    V^a(z,y) := \gamma\, \delta_{az}\qquad \textrm{for some } \gamma>0\,,
\end{equation}
which employs only one positive constant.
To see that the corresponding functions 
\begin{equation}\label{eq:WM}
W^{\alpha\beta}(z,y) := \gamma\langle L_\alpha, P_z\rangle\langle P_z,L_\beta\rangle\,,
\end{equation} 
satisfy the two required positive semi-definiteness conditions, first observe that for every $z\in Z$,
$\langle L_\alpha, P_z\rangle\langle P_z,L_\beta\rangle$ 
is a positive semi-definite matrix in $\alpha\beta$, as one easily sees by rewriting its characteristic equation\\[-4pt] 
$$\det_{\alpha,\beta}\left[\langle L_\alpha, P_z\rangle\langle P_z,L_\beta\rangle - \lambda \delta_{\alpha\beta}\right]=0$$
as the characteristic equation 
$\det\left[|P_z\rangle\langle P_z| -\lambda \textrm{id}_{L(\H)} \right]=0$ for the superoperator $|P_z\rangle\langle P_z|$ and noting that the latter has the eigenbasis $|P_1\rangle, \dots, |P_d\rangle$, whence $\lambda$ is $1$ or $0$, but in any case non-negative. But then also the upper left submatrix $\langle L_a, P_z\rangle\langle P_z,L_b\rangle$ is positive definite. Thus both (P1) and (P2) are satisfied. 

The resulting dynamics is exactly solvable. Insertion of (\ref{eq:WM}) into (\ref{eqn:Poulin}) yields the system
\begin{equation}\label{eqn:simplemod}
    \dot \rho(z,t)=\gamma\left( P_z\hat\rho(t)P_z - \rho(z,t)\right)\,,
\end{equation}
of $d$ coupled linear ODEs,
which is readily solved by first observing that 
$$P_u\hat\rho(t)P_u = P_u\hat\rho(0)P_u \textrm{ for all } u\in Z\,,$$
was one finds by summing (\ref{eqn:simplemod})
over all $z\in Z$ and then multiplying from both sides by $P_u$. This  reduces (\ref{eqn:simplemod}) to the $d$ decoupled inhomogeneous linear ODEs
$$\dot\rho(z,t) + \gamma \rho(z,t) = \gamma P_z \hat\rho(0)P_z\,.$$ 
The solution for all $z\in Z$, 
$$\rho(z,t)=e^{-\gamma t} \rho(z,0) + (1-e^{-\gamma t})P_z\hat\rho(0)P_z\,,$$
reveals exponentially fast convergence of 
$$p(z,t) = e^{-\gamma t} p(z,0) + (1-e^{-\gamma t})\tr(P_z\hat\rho(0))$$
to the axiomatically postulated probability $\tr(P_z\hat\rho(0))$ to obtain measurement outcome $m_z$
and of 
$$\hat\rho(t)=e^{-\gamma t} \hat\rho(0) + (1-e^{-\gamma t})\sum_z P_z\hat\rho(0)P_z$$
to the axiomatically postulated a priori prediction of the final (mixed) state $\sum_z P_z \hat\rho(0)P_z$.

\section{Application: Non-inertial measurement}
The dynamical implementation of projective measurements reproduces, as we saw in the previous section, precisely the measurement axioms of quantum mechanics for an inertial measurement apparatus. For the finite-dimensional quantum systems discussed here, the only non-inertial motions for a measurement apparatus would be effected by prescription of a temporally changing spatial orientation of the apparatus in terms of a time-dependent unit vector $\vec{n}(t)$ in Euclidean three-space. 

\begin{figure}[t!]
\begin{center}
\includegraphics[width=8.4cm]{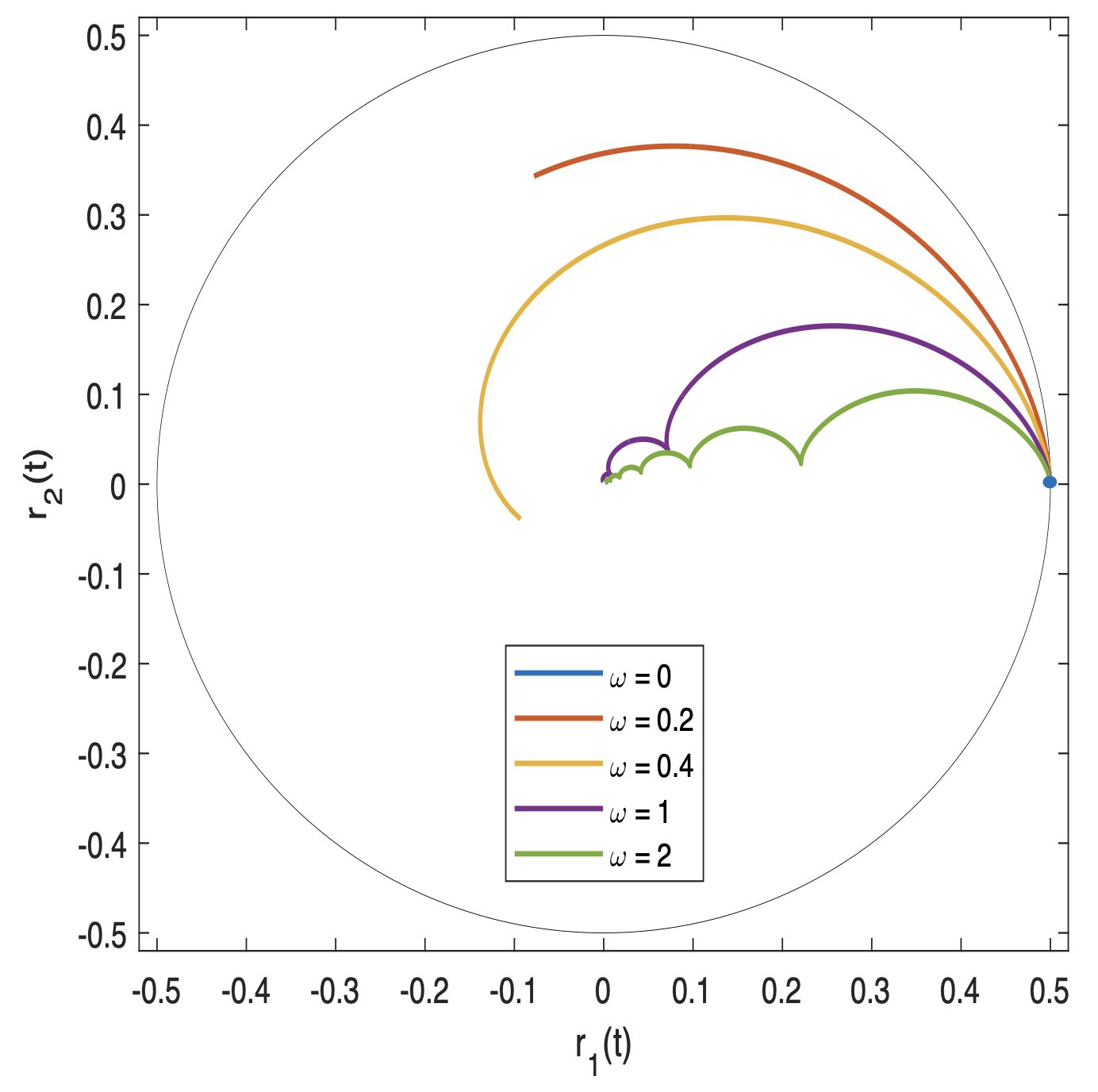}    
\caption{Temporal evolution of the Bloch vector $\vec{r}$ of a qubit in the pure initial state $\hat\rho(0)$ corresponding to $\vec{r}(0)=(1/2,0,0)$ at the beginning of the non-inertial dynamically implemented measurement specified by (\ref{eqn:rotM}) and (\ref{eqn:rotn}), for various angular velocities $\omega$ and the choice $\gamma=1$ for the numerical solution. Since $r_3(t)=0$ throughout, the entire evolution takes place in the shown equatorial plane of the Bloch ball.  }
\label{fig:Blochtrajectory}
\end{center}
\end{figure}

As a prototypical example, we calculate the predictions of the dynamical measurement implementation (\ref{eq:VM}) for a qubit  measurement apparatus with some prescribed time-dependent orientation $\vec{n}(t)$, as  described by
\begin{equation}\label{eqn:rotM}
M(t)=\sum_{m=1}^3 n^m(t) \sigma_m
\end{equation}
in terms of Pauli operators $\sigma_1,\sigma_2,\sigma_3$, which spectrally decomposes as $M(t) = P_1(t) - P_2(t)$ in terms of the orthogonal hermitian projectors 
$$P_z=\frac{1}{2}\left((-1)^{z+1} \sum_{m=1}^3 n^m(t) \sigma_m + \sigma_0\right)\quad\textrm{for }z=1,2\,,$$
where $\sigma_0$ is the identity operator on the two-dimensional Hilbert space for a qubit. The non-inertial measurement dynamics are thus described by the Diósi-Poulin equation
\begin{equation}\label{eqn:noninertialPoulin}
    \frac{1}{\gamma}\dot \rho(z,t)= P_z(t)\hat\rho(t)P_z(t) - \hat\rho(z,t)\,,
\end{equation}
which yields the equation of motion 
\begin{equation}\label{eqn:nirhohat}
\frac{1}{\gamma} \dot{\hat\rho}(t) = \sum_{z=1}^2 P_z(t) \hat\rho(t) P_z(t)  \hat\rho(t)
\end{equation}
for the quantum state $\hat\rho(t)$ induced by $\rho(z,t)$ and the equation of motion
\begin{equation}\label{eqn:niP}
    \frac{1}{\gamma} \dot p(z,t) - p(z,t) = \tr\left(P_z(t) \hat\rho(t)\right)
\end{equation}
for the induced classical probabilities $p(z,t)$. Parametrising the qubit quantum state as ${\hat\rho(t)=\frac{1}{2}\sigma_0+\sum_{m=1}^3 r_m(t)\sigma_m}$ in terms of a Bloch ball vector $\vec{r}(t)$ with ${||\vec{r}(t)||\leq\frac{1}{2}}$ and repeated use of the Pauli operator identity ${\{\sigma_a,\sigma_b\}=2\delta_{ab}\sigma_0}$, one finds that (\ref{eqn:nirhohat}) amounts to the linear equation of motion 
\begin{equation}\label{eqn:Blocheom}
    \frac{1}{\gamma} \dot{\vec{r}}(t) + \vec{r}(t) - (\vec{n}\!\cdot\!\vec{r})(t)\, \vec{n}(t)=0
\end{equation}
for the Bloch vector $\vec{r}(t)$ that can be solved, albeit it not in closed form, for given $\vec{n}(t)$. Similarly, using the same identity and the vanishing trace of the Pauli operators, one finds that (\ref{eqn:niP}) reduces to
\begin{equation}\label{eqn:PfromBlocheom}
     \frac{1}{\gamma} \dot p(z,t) - p(z,t) = \frac{1}{2} - (-1)^z (\vec{n}\!\cdot\! \vec{r})(t)\,,
\end{equation}
which, in turn, can be solved using the solution for (\ref{eqn:Blocheom}). 

We now consider concretely a non-inertial measurement apparatus (\ref{eqn:rotM}) rotating with constant angular velocity $\omega$ around the spatial $3$-axis, which corresponds to choosing
\begin{equation}\label{eqn:rotn}
\vec{n}(t)=(\cos(\omega t),\sin(\omega t),0)\,.
\end{equation}
Assuming, for definiteness, that right before the start of the measurement at $t=0$, the apparatus is initialized with $p(1,0)=p(2,0)$ and the initial quantum state is $\hat\rho(0)=P_1(0)$, one can then study the evolution of the quantum state $\hat\rho$ and of the probabilities $p(1,t)$ and $p(2,t)$ for the measurement to yield the value $+1$ or $-1$, respectively.

\begin{figure}[t!]
\begin{center}
\includegraphics[width=8.4cm]{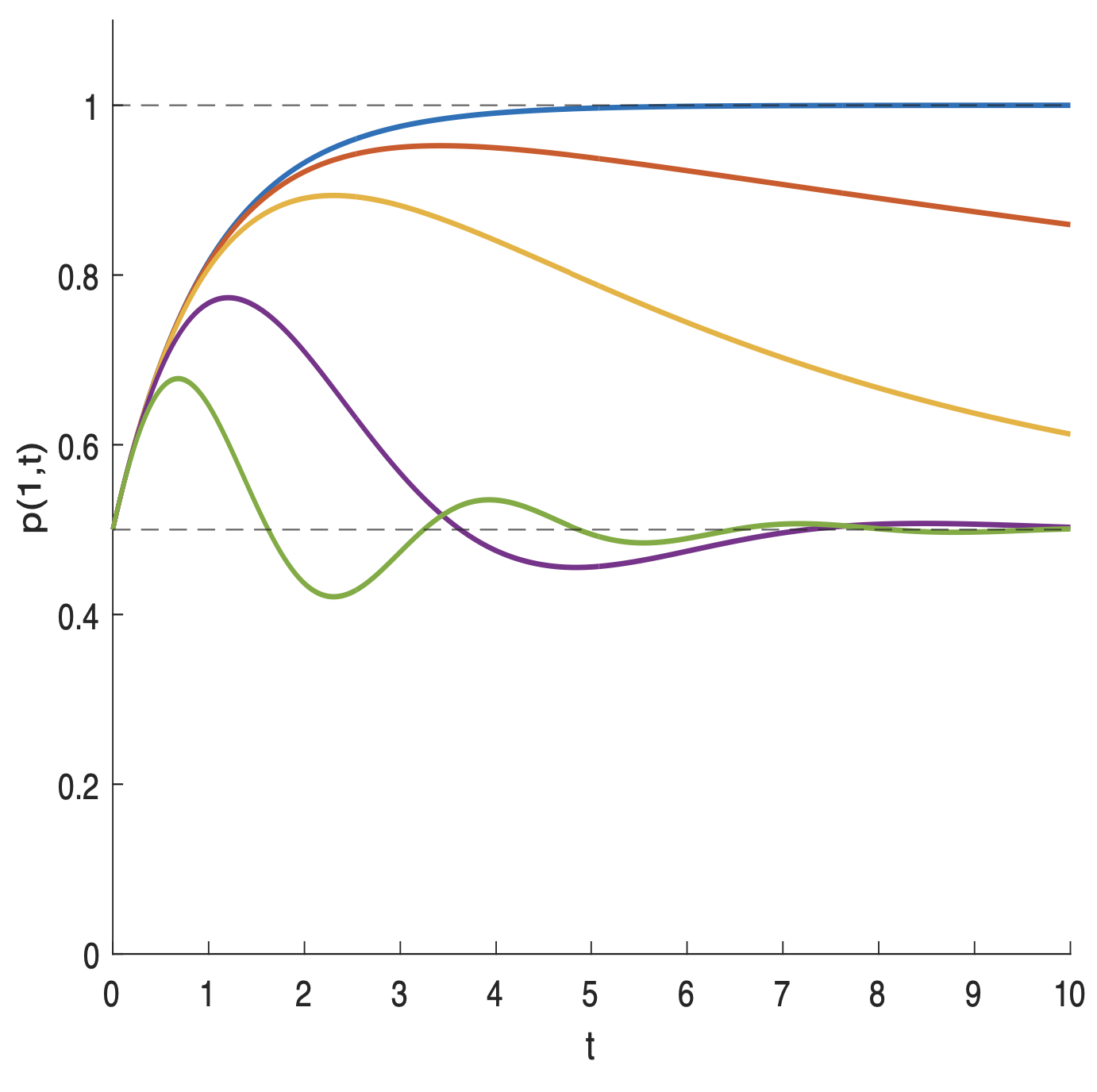}    
\caption{Probability $p(1,t)$ for dynamically modelled non-inertial measurement apparatuses, under the same conditions and with corresponding colour coding as in Fig. \ref{fig:Blochtrajectory}, to yield the measurement result $+1$ if it is read off at time $t$. While curves for $\omega$ not equal, but very close to zero (not shown), almost reach probability one, all of these finally converge to probability $0.5$ and not $1$. }
\label{fig:Probabilityconvergence.png}
\end{center}
\end{figure}

Writing the resulting equation of motion (\ref{eqn:Blocheom}) in the form $\dot{\vec{r}}(t)=-\gamma A(t) \vec{r}(t)$ for a  $3\times 3$ matrix $A(t)$ reveals that the non-equal time commutators $[A(t_1), A(t_2)]$ do not vanish for $\omega\neq 0$, which prevents an exact solution for $\vec{r}(t)$ in closed form. Readily available 
 numerical solutions, shown in Fig. \ref{fig:Blochtrajectory}, reveal that any non-inertial measurement apparatus ($\omega\neq 0$) degrades the purity of the initially pure state down to the minimum value one half, unless the dynamically implemented measurement stops before that.

 While the state evolution is experimentally not directly accessible, the probability of the measurement apparatus showing $+1$ or $-1$ at the finite time at which the measurement apparatus is read off, can be tested to arbitrary accuracy by sufficiently many repetitions. For the projective qubit measurements with only two possible outcomes, it obviously suffices to study $p(1,t)$, say. The probability of the measurement device to show the measurement value $+1$ at time $t$ is shown in Fig. \ref{fig:Probabilityconvergence.png}.

 The blue curve in Fig. \ref{fig:Probabilityconvergence.png}, for a non-rotating measurement apparatus, shows the exponentially fast convergence to the measurement result postulated by axiomatic quantum mechanics under the chosen initial conditions. But all other curves, corresponding to 
non-zero angular velocities of the measurement apparatus, finally converge to probability $0.5$ if one waits arbitrarily long. For a fixed read-off time, however, sufficiently small non-zero angular velocities yield results that are experimentally indistinguishable from an inertial measurement. 

These results are of course to be compared to the predictions of axiomatic quantum mechanics, which postulates idealised measurements that are completed instantaneously and thus are not sensitive to non-inertial motion of the measurement apparatus. Already a cursory look at real quantum measurements in the laboratory, such as a Stern-Gerlach apparatus for the measurement of the spin of charged particles, reveals that the axiomatic modelling of an instantaneous measurement is indeed an ultimately untenable idealisation. It is thus clear that measurements with sufficiently fast rotating apparatuses in the laboratory will show qualitative features as we obtained them for our dynamical implementation of projective measurements that require a finite time. 

The pursuit of combining the continuous Schrödinger-Liouville dynamics and the instantaneous projective measurement dynamics of quantum theory in one single differential equation thus forced a deviation from axiomatic quantum mechanics that implements an aspect of real measurement devices. We would have preferred to obtain an entirely equivalent reformulation of quantum theory, but the modification so inevitably forced upon us might open up new avenues that were completely barred in the standard formulation.  

At a practical modelling level, it should be interesting to determine a parametrization of the most general dynamical implementation of projective measurements that is admitted by conditions (\ref{eqn:cond1}), (\ref{eqn:cond3}), (\ref{eqn:cond2}). These conditions are derived, after all, as compatibility requirements with what axiomatic quantum mechanics requires for measurements as well.

At a more foundational level, the reformulation of previously instantaneous measurement dynamics as a now continuous evolution opens up the possibility to apply standard techniques from control theory and port-Hamiltonian systems theory to quantum systems in both the absence and the presence of measurements. 

\begin{ack}
KK and FPS acknowledge funding from the European Union Horizon Europe MSCA Grant No. 101073558 (ModConFlex). The authors would like to thank Lajos Diósi and the anonymous referees for detailed and valuable remarks. 
\end{ack}

\bibliography{main.bib}             

\end{document}